\documentstyle[twoside,aps,prl,multicol,epsf]{revtex}
\newcommand\lsim{\lower 3pt\hbox{$\buildrel < \over\sim$}}
\newcommand\gsim{\lower 3pt\hbox{$\buildrel > \over\sim$}}
\newcommand\lessgtr{\lower 3 pt\hbox{$\buildrel < \over > $}}
\begin{document}
\title{Scaling properties of three-dimensional magnetohydrodynamic turbulence}
\author{Wolf--Christian M\"uller and Dieter Biskamp}
\address{Max-Planck-Institut f\"ur Plasmaphysik,\\
85748 Garching, Germany }
\draft
\maketitle
\begin{abstract}
The scaling properties of three-dimensional magnetohydrodynamic turbulence
are obtained from direct numerical simulations of decaying turbulence 
using 512$^3$ modes. The results indicate that the turbulence does not follow
the Iroshnikov-Kraichnan phenomenology. The spectrum is consistent with $k^{-5/3}$.
In the case of hyperresistivity the structure functions exhibit a clear scaling range
yielding absolute values of the scaling exponents $\zeta_p$, in particular $\zeta_3
\simeq 1$, consistent with a recent analytical result. The scaling exponents
agree with a modified She-Leveque model $\zeta_p=p/9+1-(1/3)^{p/3}$,
corresponding to Kolmogorov scaling but sheet-like geometry of the dissipative
structures.

\end{abstract}


\begin{multicols}{2}

Magnetic turbulence is the natural state of a plasma in motion, especially
in astrophysical systems. 
The convenient framework to describe such turbulence is magnetohydrodynamics (MHD).
For high magnetic Reynolds number $Rm = vl_0/\eta$, where $v$ is a typical turbulent
velocity, $l_0$ the integral scale and $\eta$ the magnetic diffusivity,
there is a broad range of scales $l$ between $l_0$, and the dissipative scale length
$l_d$, $l_0 \gg l \gg l_d$, called the inertial range which exhibits characteristic
self-similarity or scaling properties.

The concept of inertial-range scaling was introduced by Kolmogorov \cite{1}
for hydrodynamic turbulence, which is called the Kolmogorov (K41) phenomenology. 
Assuming homogeneity and isotropy of the turbulence
as well as locality of the turbulent cascade process, he obtains
$\epsilon \sim (\delta v_l)^2/\tau_l = (\delta v_l)^3/l$ yielding the scaling
law $\delta v_l \sim \epsilon^{1/3}l^{1/3}$.
Here $\epsilon$ is the
energy dissipation rate and, to be specific, $\delta v_l = [{\bf v}({\bf x}+{\bf l})
- {\bf v} ({\bf x})]\cdot{\bf l}/l$ is the longitudinal velocity increment.
A direct consequence is the Kolmogorov energy spectrum $E_k\sim \epsilon^{2/3}
k^{-5/3}$.

For MHD turbulence the Iroshnikov-Kraichnan (IK) phenomenology \cite{2}, \cite{3}
takes into account the Alfv\'en effect, the coupling of small-scale velocity
and magnetic fluctuations by the integral-scale field $B_0$.
Hence the natural variables are the Els\"asser fields ${\bf z}^{\pm}
= {\bf v} \pm {\bf B}$, which describe Alfv\'en waves.
In the IK phenomenology the spectral transfer is reduced by the factor
$\tau_A/\tau_l$, $\epsilon \sim (\tau_A/\tau_l)(\delta z_l)^2/\tau_l$, 
where $\tau_A=l/v_A$, $v_A$ = Alfv\'en velocity in the field $B_0$,
$\delta z_l\sim \delta v_l
\sim \delta B_l \sim (\epsilon v_A)^{1/4}l^{1/4}$, and the IK energy spectrum
becomes $E_k \sim (\epsilon v_A)^{1/2}k^{-3/2}$. This  spectrum can also be
written in the form $E_k\sim \epsilon^{2/3}k^{-5/3}(kl_0)^{1/6}$ with the
integral scale $l_0$ defined by $l_0 =v_A^3/\epsilon$, which illustrates
the nonlocal character of the energy cascade in MHD turbulence. 

It is, however, well known that these qualitative scaling relations for $\delta v_l$
or $\delta z_l$ are not exactly valid in a statistical sense because of
intermittency, which implies that the distribution of turbulent scales is not 
strictly self-similar. A quantitative measure is provided by the scaling
exponents $\zeta_p$ of the structure functions, 
the moments of the field increments.
For hydrodynamic turbulence
She and Leveque \cite{4} proposed a model leading to the expression 
$\zeta^{\rm SL}_p = p/9 +2[1-(2/3)^{p/3}]$, which fits the experimental results
surprisingly well, reproducing in particular the exact result $\zeta_3=1$. 
This model has been modified for MHD incorporating the IK effect
\cite{5}, \cite{6}, which yields $\zeta^{\rm IK}_p = p/8 + 1 -(1/2)^{p/4}$,
in particular $\zeta^{\rm IK}_4 =1$. 
 
The IK phenomenology has been supported by direct numerical simulations
of 2D MHD turbulence at moderate Reynolds numbers \cite{7}.
However, recent developments in MHD turbulence theory cast some doubt
on the general validity of the IK scaling. 2D 
simulations at considerably higher Reynolds numbers reveal an anomalous
scaling behavior \cite{8}, \cite{9}, indicating that the results of Ref.\ \cite{7}
are not asymptotic. There have also been theoretical arguments in favor of a
Kolmogorov scaling, e.g., \cite{9a}, \cite{9b}. Even more convincingly,
exact relations have
been derived for moments of certain triple products of $\delta z_l$ \cite{10},  
which are shown to be
proportional to $l$, i.e., $\zeta_3 =1$, analogous to the well-known 4/5-relation
in hydrodynamic turbulence, thus excluding the IK result $\zeta_4 =1$.
Scaling exponents for MHD turbulence have also been derived from observations
in the solar wind \cite{12}. Here agreement with the IK exponents
has been claimed \cite{5}, but in this comparison the observational results were 
normalized {\it assuming} $\zeta_4=1$. Actually
the error bars seem to be  too large to reach a definite conclusion.

To clarify the issue of scaling in 3D MHD turbulence
direct numerical simulations are desirable
with higher Reynolds numbers than studied previously, for instance in \cite{12a},
\cite{12b}, \cite{12c},
\cite{12d}, \cite{12e}.
In this Letter we present a numerical study of freely decaying
turbulence with spatial resolution of $512^3$ modes. 
The scaling properties are analyzed by considering the time-averages of the normalized
spectra and structure functions. 
We solve the incompressible MHD equations
\begin{equation}
\partial_t{\bf B} -\nabla\times({\bf v}\times{\bf B})=\eta_{\nu}(-1)^{\nu-1}
\nabla^{2\nu}{\bf B},
\label{1}
\end{equation}
\begin{equation}
\partial_t{\bf w} -\nabla\times({\bf v}\times{\bf w})-\nabla \times({\bf j}\times
{\bf B})=\mu_{\nu}(-1)^{\nu-1}\nabla^{2\nu}{\bf w},
\label{2}
\end{equation}
\[
{\bf w} = \nabla \times {\bf v}, \quad {\bf j}=\nabla\times{\bf B},\quad \nabla\cdot
{\bf v}=\nabla\cdot{\bf B}=0,
\]
by applying a pseudo-spectral method with spherical mode truncation as conveniently
used in 3D turbulence simulations (instead of full dealiasing by the 2/3 rule
used in most 2D simulations). The generalized magnetic Prandtl number
$\eta_{\nu}/\mu_{\nu}$ has been set equal to unity. Initial conditions are
\begin{equation}
{\bf B}_{\bf k} = a\, {\rm e}^{-k^2/k^2_0-i\alpha_{\bf k}},
\quad {\bf v}_{\bf k} = b\, {\rm e}^{-k^2/k^2_0-i\beta_{\bf k}},
\label{3}
\end{equation}
which are characterized by random phases $\alpha_{\bf k}$, $\beta_{\bf k}$ and
satisfy the conditions ${\bf k}\cdot{\bf B}_{\bf k}={\bf k}\cdot{\bf v}_{\bf k}=0$,
$E=E^V +E^M=1$ and $E^V/E^M=1$. Further restrictions on ${\bf B}_{\bf k}$ 
arise by requiring a specific value of the magnetic helicity 
$H = \int d^3x\, {\bf A}\cdot{\bf B}$. We believe that 
finite magnetic helicity is more typical than $H\simeq 0$, since 
MHD turbulence usually occurs in rotating systems.
The wavenumber $k_0$, the location of the maximum of the initial energy spectrum,
is chosen $k_0=4$, which allows the inverse cascade of $H_{\bf k}$ to develop
freely during the simulation time of 10 time units (about 7 eddy turnover
times, defining the eddy turnover time as the time required to reach the maximum
dissipation from the smooth initial state). Though this choice implies 
a certain loss of inertial range,
the sacrifice is unavoidable in the presence of inverse cascade dynamics, since assuming
$k_0\sim 1$ would lead to magnetic condensation in the lowest-$k$ state,
which would also affect the dynamics of higher-$k$ modes.
Both normal diffusion $\nu=1$ and hyperdiffusion $\nu=2$ have been used,
$\nu =1$ to discuss the spectral properties  and $\nu=2$ to determine the scaling of
the structure functions. 
All runs presented in this Letter have finite $H$, $H/H_{\max} \simeq 0.6$,
and negligible alignment. 
Table \ref{t1} lists the important parameters of the simulation runs, where
the magnetic Taylor Reynolds number is $Rm_{\lambda}= Rm^{1/2}$. 
Since $Rm_{\lambda}$ is not stricly constant
during turbulence decay but increases slowly $Rm_{\lambda}\sim t^{1/8}$, 
we give the values taken  at a specific time $t=4$.
 
\begin{figure}
\epsfxsize=9truecm
\epsfbox{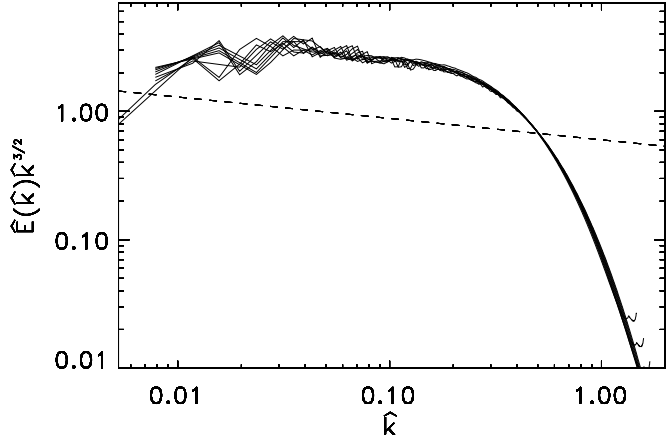}
\caption{Scatter plot of the normalized angle-integrated energy spectrum 
compensated with $k^{3/2}$ from run 3. The dashed line indicates the $k^{-5/3}$
spectrum.}  
\label{f1}
\end{figure}

\begin{figure}
\epsfxsize=9truecm
\epsfbox{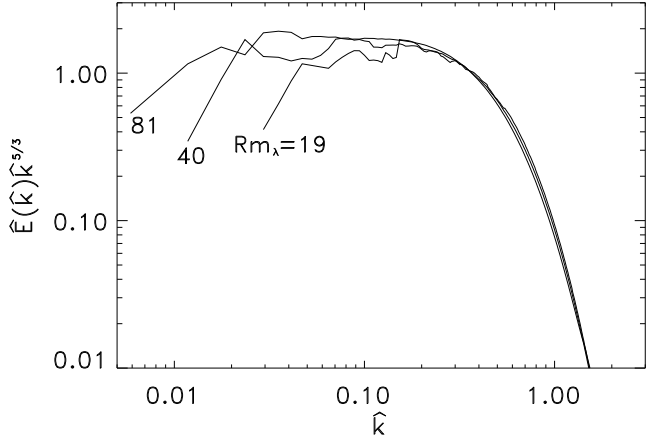}
\caption{Time-averaged normalized energy spectra compensated with $k^{5/3}$ from runs
1-3.}
\label{f2}
\end{figure}

We first discuss the spectral properties considering the angle-averaged
energy spectrum $E_k$. Figure \ref{f1} shows the scatter plot
of the normalized spectrum (the normalization is discussed below), 
compensated by $k^{3/2}$, taken from run 3 over the period $t=4-10$ of fully developed
turbulence. The spectrum exhibits a clear scaling range of almost one 
decade with a spectral law, which is definitely steeper than the IK spectrum $k^{-3/2}$,
 close to (in fact slightly steeper than) 
$k^{-5/3}$ indicated by the dashed
line.
In order to form the time average in a system of decaying turbulence,
the spectrum must be normalized to eliminate the time variation of the macroscopic
quantities. In hydrodynamic turbulence the only such quantity is $\epsilon$,
which leads to the universal form of the Kolmogorov spectrum
$\widehat{E}(\widehat{k})= 
E_k/(\epsilon \eta^5)^{1/4}$, $\widehat{k}=kl_K$, where $l_K=(\eta^3/\epsilon)^{1/4}$ 
is the Kolmogorov
length. However, when normalized in this way the MHD energy spectrum
is found to change during turbulence decay and
even more strongly so when comparing runs of different $Rm_{\lambda}$. For
finite magnetic helicity the spectrum may also depend on $H$, which introduces
a second macroscopic length scale $l_1=H/v^2_A$ in addition to $l_0$, i.e,
the spectrum may contain some function of $l_0/l_1$. To determine this function 
we propose the following argument.
Since the Alfv\'en effect is clearly present in the simulations,
kinetic and magnetic energy spectrum being nearly equal at small scales, 
while on the other hand the scaling $E_k \sim k^{-5/3}$ is observed,     
we modify the nonlinear 
transfer in the IK ansatz by a factor $(l/l_1)^{\delta}$,
\begin{equation}
\frac{\tau_A}{\tau_l}\left(\frac{l}{l_1}\right)^{\delta}\frac{\delta z_l^2}{\tau_l}
=\epsilon,
\label{4}
\end{equation}
and determine $\delta$ by
requiring the observed scaling $\delta z_l \sim l^{1/3}$, which gives  
$\delta =-1/3$ and hence 
\begin{equation}
E_k \sim \epsilon^{2/3}(l_0/l_1)^{1/6} k^{-5/3}.
\label{5}
\end{equation} 
Also the dissipation scale length is slightly changed.
Balancing nonlinear transfer and dissipation gives
\begin{equation}
l^H_K =
 l_K (l_0/l_1)^{1/8}.
\label{6}
\end{equation}
Using these relations we obtain a  the normalized energy spectrum
$\widehat{E}(\widehat{k})=E_k/[(\epsilon\eta^5)^{1/4}(l_0/l_1)^{3/8}]
=\widehat{k}^{-5/3}F(\widehat{k})$, $\widehat{k}=kl^H_K$. 
Normalized in this way the spectra  at different times of run 3
coincide very well as seen in Fig.\ \ref{f1} and so do the time-averaged
normalized energy spectra of runs 1-3
shown in Fig.\ \ref{f2}, which vary only by the extent of the inertial
range, apart from statistical oscillations. 
Relations (\ref{5}), (\ref{6}) are not valid for $H\simeq 0$,
where we expect the pure Kolmogorov normalization to be valid.

A more complete picture of the inertial-range distribution of turbulent structures is provided
by the scaling exponents $\zeta_p$ of the structure functions, where the
second order exponent $\zeta_2$ is related to the inertial-range spectral law $\sim
k^{-(1+\zeta_2)}$. To be definite we consider the moments of the absolute value
of the longitudinal increments $\delta z^{\pm}_l$ discussing only 
the runs 4 and 5 with the highest Reynolds numbers. The normalized structure
functions $\widehat{S}^{\pm}_p(\widehat{l}) = \langle|\delta z^{\pm}_l|^p\rangle/E^{p/2}$,
$\widehat{l}=l/l^H_K$, are averaged over time, exhibiting a similar weak scatter
as for the spectrum in Fig.\ \ref{f1}. 

\begin{figure}
\epsfxsize=9truecm
\epsfbox{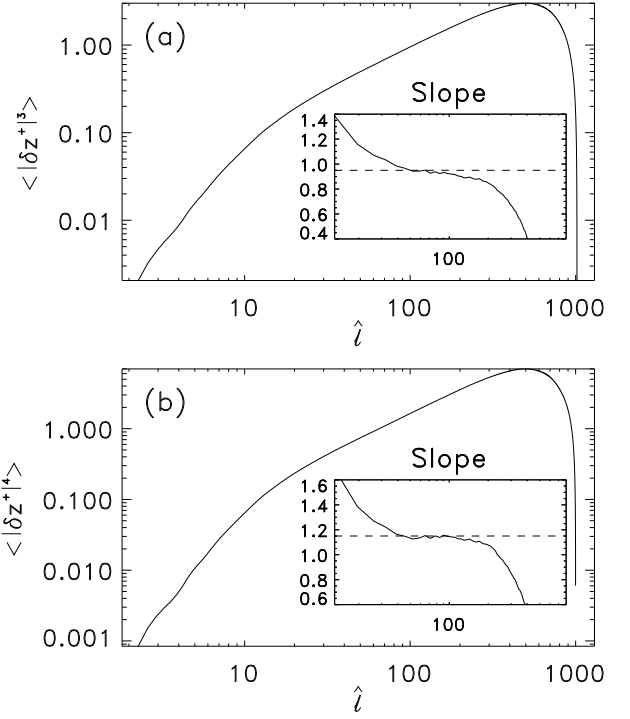}
\caption{Log-log plot of the time-averaged normalized structure functions from
run 4. (a) $\widehat{S}^+_3(\widehat{l})$,
(b) $\widehat{S}^+_4(\widehat{l})$. The inserts give the derivatives, the
horizontal dashed lines provide most probable values of the scaling exponents.}
\label{f3}
\end{figure}

For normal diffusion $\nu=1$ no  scaling range is visible. (Note that
the structure function $S_2$ corresponds to the one-dimensional spectrum
$E_{k_x}$, which has a shorter inertial range than the angle-averaged spectrum shown
in Fig.\ \ref{f1}.) For $\nu=2$, however, there is a scaling range $30<\widehat{l}
<200$, as seen in Fig.\ \ref{f3}, where the time-averaged curves 
$\widehat{S}^+_p$ are plotted for $p=3,4$. The inserts give the logarithmic
derivatives, where the central quasi-constant parts determine the scaling coefficients,
the dashed horizontal lines indicating the most probable values
$\zeta^+_3 \simeq 0.95$, $\zeta^+_4 \simeq 1.15$. These results
are consistent with the spectral law derived from Fig.\ \ref{f1} and are close to
the analytical prediction $\zeta_3=1$. It is true that the analytical theory
refers to third-order moments different from $\langle|\delta z|^p\rangle$ 
discussed here, but the scaling coefficients should not depend thereof (the
scaling range, however, does). One might object that the use of hyperdiffusion
affects the inertial-range scaling, if the scaling range is not very broad.
In fact, the energy spectrum law tends to be polluted by the bottleneck
effect, which is particularly pronounced for hyperdiffusion (see e.g., \cite{13}).
Thus the energy spectrum in run 4 (not shown) is effectively flatter than $k^{-5/3}$
expected from the value of $\zeta_2$. However, there is, to our knowledge, no
argument for a similar  effect in the structure functions.  
 
\begin{figure}
\epsfxsize=9truecm
\epsfbox{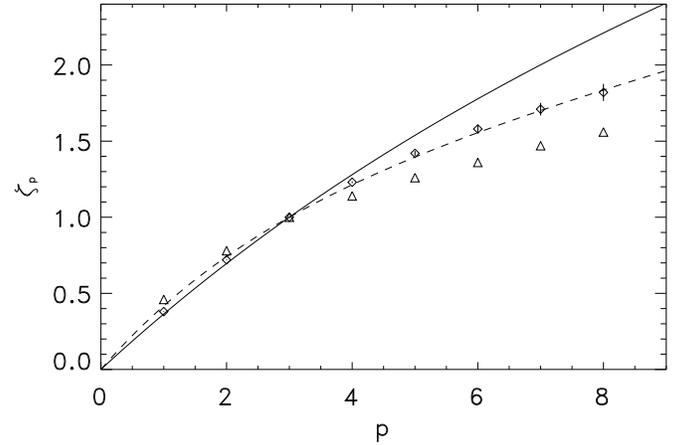}
\caption{Scaling exponents $\zeta^+_p$ for 3D MHD
turbulence (diamonds) and relative exponents $\zeta^+_p/\zeta^+_3$ for 
2D MHD turbulence (triangles). The continuous curve is the She-Leveque
model $\zeta^{\rm SL}_p$, the dashed curve the modified model $\zeta^{\rm MHD}_p$.}
\label{f4}
\end{figure}

Assuming the exact result $\zeta_3=1$ allows to obtain rather accurate values of $\zeta_p$ by using
the property of ESS (extended self-similarity) 
\cite{14} plotting $S_p$ as function of $S_3$. (It should be noted that 
ESS usually results in almost perfect scaling behavior, but the scaling
coefficients thus derived vary in time, hence time averaging is required.)
The results are shown in Fig.\ \ref{f4}, which gives the ESS results of
$\zeta^+_p$ for run 4 (diamonds) and the ESS values $\xi^+_{3p} =\zeta^+_p/\zeta^+_3$
from \cite{15} for 2D MHD turbulence (triangles). (In 2D MHD the absolute 
values of $\zeta_p$ are found in \cite{9} to decrease with $Rm$, while
the relative values $\xi_{3p}$ appear to be independent of $Rm$.)
The results indicate that in 3D MHD turbulence is less intermittent than 2D,
but it is more intermittent than hydrodynamic turbulence,
the continuous curve, which gives the She-Leveque result $\zeta^{\rm SL}_p$. 
As shown by Politano and Pouquet \cite{6},
the She-Leveque concept contains effectively three parameters:
$g$ related to the scaling $\delta z_l\sim l^{1/g}$, $x$ related to the
energy transfer time at the smallest, the dissipative scales $t_l \sim l^x$, and
$C$, the codimension of the dissipative structures,
\begin{equation}
\zeta_p = \frac{p}{g}(1-x) +C\left(1-(1-x/C)^{p/g}\right).
\label{7}
\end{equation}
Our results for the 3D MHD case suggest Kolmogorov scaling $g=3$, $x=2/3$,
while different from hydrodynamic turbulence the dissipative structures
are sheet-like, hence the codimension is $C=1$,
\begin{equation}
\zeta^{\rm MHD}_p = p/9 + 1-(1/3)^{p/3}.
\label{8}
\end{equation}
This is the dashed curve in Fig.\ \ref{f4}, which fits the numerical values
very well. 

In conclusion we have studied the spatial scaling properties of
3D MHD turbulence using direct numerical simulations with resolution
of $512^3$ modes. The results indicate that the turbulence does not
follow the Iroshnikov-Kraichnan (IK) phenomenology. The energy spectrum 
is consistent with a $k^{-5/3}$ law. For hyperresistivity the structure functions
exhibit a clear scaling range yielding absolute values of the scaling
exponents $\zeta_p$, in particular $\zeta_3\simeq 1$, consistent with
recent analytical predictions. The scaling exponents agree well with
a modified She-Leveque model $\zeta^{\rm MHD}_p=p/9+1-(1/3)^{p/3}$, corresponding
to Kolmogorov scaling, but sheet-like geometry of the dissipative structures.
The results are also consistent with observations of turbulence in the
solar wind, which typically show a $k^{-1.7}$ spectrum. 

The authors would like to thank 
Andreas Zeiler for  providing the basic version of the code, Antonio Celani
for developing some of the diagnostics, 
and Reinhard Tisma for
optimizing the code for the CRAY T3E.

\newpage

\narrowtext
\begin{table}
\caption{Summary of the simulation runs.} 
\begin{center}
\begin{tabular}{ccccc}
run No  &  $N$  &   $\nu$  &  $\eta_{\nu}$ & $Rm_{\lambda}$  \\ \hline
   1  &     128   &     1   &   $3\times 10^{-3}$ & 19           \\
   2  &     256   &     1   &     $10^{-3}$ &        40          \\
   3  &     512   &     1   &   $3\times 10^{-4}$ &    81         \\
   4 &     512   &     2   &   $3\times 10^{-8}$ &     --       \\
\end{tabular}
\end{center}
\label{t1}
\end{table}

\narrowtext
\narrowtext
\narrowtext

\end{multicols}

\begin{references}
\bibitem{1}
A.\ Kolmogorov, Dokl.\ Akad.\ Nauk SSSR {\bf 31}, 538 {1941}. 
\bibitem{2}
P.\ S.\ Iroshnikov, Astron.\ Zh. {\bf 40}, 742 (1963), [Sov.\ Astron. {\bf 7},
568 (1964)]. 
\bibitem{3}
R.\ H.\ Kraichnan, Phys.\ Fluids {\bf 8}, 1385 (1965).
\bibitem{4}
Zh.-S.\ She and E.\ Leveque, Phys.\ Rev.\ Lett. {\bf 72}, 336 (1994).
\bibitem{5}
R.\ Grauer, J.\ Krug, and C.\ Marliani, Phys.\ Lett. A {\bf 195}, 335 (1994).
\bibitem{6}
H.\ Politano and A.\ Pouquet, Phys.\ Rev. E {\bf 52}, 636 (1995).
\bibitem{7} 
D.\ Biskamp and H.\ Welter, Phys.\ Fluids B {\bf 1}, 1964 (1989).
\bibitem{8}
D.\ Biskamp, E.\ Schwarz, and A.\ Celani, 
Phys.\ Rev.\ Lett. {\bf 81}, 4855 (1998).
\bibitem{9}
D.\ Biskamp and E.\ Schwarz, to be published.
\bibitem{9a}
M.\ Verma, M.\ L.\ Goldstein, S.\ Gosh, and W.\ T.\ Stribling,
J.\ Geophys.\ Res. {\bf 101}, 21619 (1996).
\bibitem{9b}
P.\ Goldreich and S.\ Sridhar, Astrophys.\ J. {\bf 485}, 680 (1997).
\bibitem{10}
H.\ Politano and A.\ Pouquet, Phys.\ Rev. E {\bf 57}, R21 (1998), and
Geophys.\ Res.\ Lett. {\bf 25}, 273 (1998).
\bibitem{12}
L.\ F.\ Burlaga, J.\ Geophys.\ Res. {\bf 96}, 5847 (1991).
\bibitem{12a}
A.\ Pouquet, M.\ Meneguzzi, and U.\ Frisch, Phys.\ Rev. A {\bf 33}, 4266 (1986).
\bibitem{12b}
S.\ Kida, S.\ Yanase, and J.\ Mizushima, Phys.\ Fluids A {\bf 3}, 457 (1991).
\bibitem{12c}
M.\ Hossain, P.\ C.\ Gary, D.\ H.\ Pontius, and W.\ H.\ Matthaeus,
Phys.\ Fluids {\bf 7}, 2886 (1995).
\bibitem{12d}
H.\ Politano, A.\ Pouquet, and P.\ L.\ Sulem, Phys.\ Plasmas {\bf 2}, 2931 (1995).
\bibitem{12e}
A.\ Brandenburg, R.\ J.\ Jennings, A.\ Nordlund, M.\ Rieutord, R.\ F.\ Stein, and
I.\ Tuominen, J.\ Fluid Mech. {\bf 306}, 325 (1996). 
\bibitem{13}
V.\ Borue and  S.\ A.\ Orszag, Europhys.\ Lett. {\bf 29}, 687 (1995).
\bibitem{14}
R.\ Benzi, S.\ Ciliberto, R.\ Tripiccione, C.\ Baudet, F.\ Massaioli, and S.\ Succi,
Phys.\ Rev. E {\bf 48}, R29 (1993).
\bibitem{15}
H.\ Politano, A.\ Pouquet, and V.\ Carbone, Europhys.\ Lett. {\bf 43}, 516 (1998).
\end{references}
\end{document}